\documentclass[conference]{IEEEtran}
\IEEEoverridecommandlockouts
\usepackage{cite}
\usepackage{amsmath,amssymb,amsfonts}
\usepackage{graphicx}
\usepackage{textcomp}
\usepackage{xcolor}
\usepackage[dvipsnames]{xcolor}
\usepackage{wasysym}
\usepackage{graphicx}
\usepackage{pifont}
\newcommand{\cmark}{\ding{51}}%
\newcommand{\xmark}{\ding{55}}%
\usepackage{algorithm}
\usepackage{algpseudocode}
\usepackage{float}
\usepackage{tabularx}
\usepackage{booktabs}
\usepackage{array}
\usepackage{subcaption}
\usepackage{makecell}

\usepackage{multirow}
\usepackage{booktabs}

\renewcommand{\thesection}{\arabic{section}}
\renewcommand{\thesubsection}{\thesection.\arabic{subsection}}

\newcommand{\PhaseComment}[1]{\Statex \hspace{-\algorithmicindent}\textbf{#1}}

\def\BibTeX{{\rm B\kern-.05em{\sc i\kern-.025em b}\kern-.08em
    T\kern-.1667em\lower.7ex\hbox{E}\kern-.125emX}}

\begin{document}




\title{\textsc{Dart}: A DAG-Based Reputation and Incentive Framework via Blockchain-Enabled Governance for Trustworthy LLM Multi-Agent Collaboration}



\author{
\IEEEauthorblockN{Manoj Kumal${^a}$, Xinyun Liu${^a}$, Ronghua Xu${^a}$\IEEEauthorrefmark{1}}

\IEEEauthorblockA{
$^{a}$Department of Applied Computing, Michigan Technological University, Houghton, MI 49931, USA\\ 
\{mkumal, xinyunl, ronghuax\}@mtu.edu
}
}

\maketitle

\begin{abstract}

Large language model (LLM)-based multi-agent systems (MAS) predominantly rely on centralized orchestration and lack formal verification mechanisms for agent reliability, participation, and system-level behavioral alignment.
These shortcomings leave open environments severely vulnerable to uncooperative or malicious agents.
This work proposes \textsc{Dart}, a Directed Acyclic Graph (DAG)-based 
reputation and incentive regulation framework for trustworthy multi-agent collaboration, combining centralized operational orchestration with blockchain-enabled decentralized governance and accountability.
\textsc{Dart} unifies DAG workflow orchestration, capability and reputation-aware task allocation, dynamic behavior updates, multi-factor incentives, and smart contract accountability paired with IPFS storage.
Under this paradigm, agent selection dynamically balances task alignment, historical reputation, and workload, while post-execution behavioral evidence continuously calibrates agent trust and the probability of future participation.
Evaluated across four axes, \textsc{Dart} achieves 93.6\% Pass@1 on GSM8K and builds a full-stack application in 142 s using two agents, 
outperforming centralized baselines.
Across five independent 150-round longitudinal trials, Full \textsc{DART} achieves a mean task success rate of 93.33 ± 2.26\%, output quality of 0.9357 ± 0.0117, retry rate of 0.2307 ± 0.0816, and allocation delay of 1.1153 ± 0.0408 s, consistently outperforming its ablated configurations
\textsc{Dart} isolates persistent and intermittent malicious agents, obtaining a 99.3\% output containment rate and restoring system success to 99.8\%.
These results demonstrate the potential of coupling reputation, incentives, DAG-based coordination, and verifiable blockchain-enabled governance to support adaptive and accountable multi-agent collaboration. 


\end{abstract}


\section{Introduction}
The rapid proliferation of large language models (LLMs) and vision-language models (VLMs) has demonstrated remarkable capabilities in human intent understanding, complex reasoning, and task planning~\cite{wang2025history, yao2022react}.
Building on these advances, LLM-driven multi-agent systems (MAS) have evolved beyond isolated, task-specific models into dynamic ecosystems of autonomous, interactive AI agents~\cite{wang2025internet}.
Whereas traditional MAS rely on static, predefined protocols or centralized controllers to orchestrate distributed intelligence~\cite{guo2024large}, modern LLM agents integrate multimodal perception, reasoning, memory, and tool utilization to execute goal-directed behavior with minimal human oversight \cite{liu2026security}.
Consequently, these LLM-based agentic systems are increasingly deployed across diverse application domains, including scientific discovery~\cite{schmidgall2025agent}, vulnerability detection~\cite{widyasari2025let}, robotics~\cite{chen2025multi}, healthcare~\cite{borkowski2025multiagent}, and complex problem-solving~\cite{chen2024comm}.

Recent research has introduced seminal LLM-driven multi-agent frameworks such as CAMEL~\cite{li2023camel}, MetaGPT~\cite{hong2024metagpt}, AgentVerse~\cite{chen2024agentverse}, ChatDev~\cite{qian2024chatdev}, and MegaAgent~\cite{wang2025megaagent} that enable autonomous agents to communicate via natural language and execute complex, reasoning-driven collaboration. Nevertheless, a critical open question remains: How can autonomous agents be governed to ensure that their interactions are demonstrably trustworthy, verifiable, and incentive-aligned in open multi-agent environments?


Existing frameworks predominantly rely on centralized orchestration for task decomposition, assignment, and result aggregation. Consequently, they inherit classic centralized vulnerabilities, including single points of failure and severe scalability bottlenecks~\cite{ding2025decentralized}.
To mitigate these issues, decentralized multi-agent architectures integrate blockchain technology and smart contracts to enforce automated rules and maintain immutable, transparent interaction logs~\cite{karim2025ai}.
However, current blockchain-based agent frameworks~\cite{jin2024decoagent,chen2024blockagents,chong2025llm,qi2026towards} focus primarily on transactional integrity, largely neglecting intelligent reasoning and adaptive collaboration.
Furthermore, existing paradigms lack efficient, scalable mechanisms to model complex agent workflows that capture dependencies and interaction chains across multi-agent executions.
Finally, reputation and incentive mechanisms in these systems remain loosely coupled, thereby failing to tie rewards and penalties directly to the verifiable quality of task completion or the fidelity of collaborative behavior.

\begin{table}[tbp]
\centering
\small
\caption{Comparison with LLM multi-agent frameworks}
\label{tab:comparison}
\setlength{\tabcolsep}{1pt}
\begin{tabular}{lcccc}
\toprule
\textbf{Method} & 
\rotatebox{65}{\makecell[c]{\textbf{Decentralized} \\ \textbf{ Governance}}} & 
\rotatebox{65}{\makecell[c]{\textbf{Agent Behavior} \\ \textbf{Tracking}}} & 
\rotatebox{65}{\makecell[c]{\textbf{Adaptive} \\ \textbf{Coordination}}} &
\rotatebox{65}{\makecell[c]{\textbf{Incentive} \\ \textbf{Compatibility}}} \\
\midrule
CAMEL~\cite{li2023camel} & $\times$ & $\times$ & $\times$ & $\times$\\
MetaGPT~\cite{hong2024metagpt} & $\times$ & \checkmark & $\times$ & $\times$\\
AgentVerse~\cite{chen2024agentverse} & $\times$ & $\times$ & \checkmark & $\times$\\
ChatDev~\cite{qian2024chatdev} & $\times$ & \checkmark & \checkmark & $\times$\\
MegaAgent~\cite{wang2025megaagent} & $\times$ & \checkmark & \checkmark & $\times$\\
DecoAgent~\cite{jin2024decoagent} & \checkmark & \checkmark & $\times$ & \LEFTcircle\\
BlockAgents~\cite{chen2024blockagents} & \checkmark & \checkmark & \checkmark & \LEFTcircle\\
LLM-Net~\cite{chong2025llm} & \checkmark & \checkmark & \checkmark & \LEFTcircle \\
Qi et al~\cite{qi2026towards} & \checkmark & \checkmark & $\times$ & \LEFTcircle \\
\textbf{\textsc{Dart} (Our)} & \checkmark & \checkmark & \checkmark & \checkmark \\
\bottomrule
\end{tabular}
{\footnotesize
$\checkmark$\,=\,Supported\quad
$\LEFTcircle$\,=\,Partial\quad
$\times$\,=\,Not supported
}
\vspace{-10pt}
\end{table}


To bridge these critical gaps, we propose \textsc{Dart}, a novel DAG-based agent reputation framework for trustworthy LLM multi-agent systems with blockchain-enabled governance and accountability. \textsc{Dart} seamlessly unifies Directed Acyclic Graph (DAG)-driven workflow modeling, behavior-shaping incentive schemes, and smart contract-enabled validation. As summarized in Table~\ref{tab:comparison}, \textsc{Dart} addresses key structural challenges in modern multi-agent systems by integrating granular behavior tracking, adaptive coordination, incentive regulation, and blockchain-enabled accountability.

\textsc{Dart} leverages blockchain technology and smart contracts as core infrastructure to provide a tamper-resistant and verifiable foundation for agent collaboration and accountability.
We employ the InterPlanetary File System (IPFS) as off-chain storage to maintain persistent, content-addressable memory for raw execution data generated by agentic workflows. Concurrently, smart contracts form an on-chain computational layer that enforces secure and verifiable interactions, including agent, task, and action registration, as well as automated reputation score updates.
Second, \textsc{Dart} incorporates a DAG structure to model agentic workflows and govern task dependencies, scheduling, and execution. This design yields a hierarchical, collaborative coordination scheme that provides end-to-end lifecycle management spanning planning, execution, monitoring, and feedback, thereby supporting
explainable task dependencies and auditable execution traces.
Third, we introduce a behavior-shaping reputation mechanism that jointly evaluates agent performance across skill reliability, collaboration quality, and task execution metrics. By combining long-term trust derived from historical reputation scores with short-term artifacts from real-time execution data, \textsc{Dart} dynamically adjusts task assignments and performs adaptive coordination to optimize overall system output. Furthermore, a multi-factor incentive scheme aligns individual execution criteria with collective performance goals. Malicious or abnormal behavior is automatically detected, resulting in reduced assignment privileges or complete isolation to prevent cascading failures.
Ultimately, \textsc{Dart} establishes long-term agent accountability, operational reliability, and incentive-aligned collaboration through blockchain-enabled governance in multi-agent systems.

The major contributions of this research are as follows:

\begin{itemize}

    \item We propose \textsc{Dart}, a hybrid trustworthy multi-agent collaboration framework that systematically integrates DAG-based agent workflow modeling and orchestration and blockchain-enabled verifiable infrastructure atop a dual-layer architecture consisting of on-chain governance and off-chain storage.

    \item We formulate a behavior-shaping reputation that combines historical trust scores with real-time execution feedback by evaluating agents' reliability, collaborative fidelity, and task performance.

    \item We design multi-factor incentives and penalties that motivate honest agents to align individual performance with collective system goals while automatically detecting, restricting, or isolating malicious agents.

    \item We validate the effectiveness and robustness of \textsc{Dart} through extensive experiments across diverse benchmarks and heterogeneous agents, demonstrating its ability to improve the long-term performance, reliability, and behavioral accountability in multi-agent systems. 
    
    
\end{itemize}

The rest of this paper is organized as follows: Section~\ref{sec:relatedwork} reviews related work in LLM multi-agent systems, blockchain-based coordination, and behavior-shaping agent reputation. Section~\ref{sec:design} presents the system architecture and implementation details of the \textsc{Dart} framework.
Section~\ref{sec:reputation_incentive} formalizes the proposed reputation model and multi-factor incentive scheme.
Section~\ref{sec:experiments} details the experimental setup, evaluation metrics, and comparative analysis.
Section ~\ref{sec:discussion} discusses the key findings, practical implications, and limitations of the proposed framework. Finally, Section ~\ref{sec:conclusion} concludes the paper and outlines directions for future research.

\section{Related Work}
\label{sec:relatedwork}

\subsection{LLM-Based Multi-Agent Systems}

Built upon high-capacity Transformer architectures~\cite{ashish2017attention, achiam2023gpt, touvron2023llama}, Large Language Models (LLMs) possess advanced reasoning, planning, and tool utilization capabilities.
By leveraging natural language communication, LLM-driven Multi-Agent Systems (MAS) decompose complex tasks into sub-problems distributed across specialized agents, effectively mimicking human team collaboration. 
Early representative frameworks explored general-purpose coordination via agent chaining and tool-augmented reasoning—such as AutoGPT~\cite{autogpt} for autonomous task decomposition,
BabyAGI~\cite{babyagi} for dynamic prioritization,
and LangGraph~\cite{langgraph} for long-running and stateful workflow execution. 

Subsequent architectures formalized domain-specific agent roles and operational workflows. For instance, CAMEL~\cite{li2023camel}, MetaGPT~\cite{hong2024metagpt}, and ChatDev~\cite{qian2024chatdev} employ role-based planning, while AgentVerse~\cite{chen2024agentverse} and MegaAgent~\cite{wang2025megaagent} utilize expert orchestrators to coordinate software engineering pipelines.
However, frameworks such as MetaGPT~\cite{hong2024metagpt} and ChatDev~\cite{qian2024chatdev} rely on isolated single-agent planning, where a centralized agent decomposes tasks without collaborative input from domain-specific peers.
Similarly, while MegaAgent~\cite{wang2025megaagent} introduces hierarchical coordination, its top-down controller induces context fragmentation across the agent hierarchy and lacks intra-domain expert collaboration.
Furthermore, AgentVerse~\cite{chen2024agentverse} fosters expert consensus but assumes static environmental conditions, limiting its adaptability in dynamic settings. 

Most existing LLM multi-agent frameworks therefore rely on centralized or semi-centralized orchestration, creating potential single points of failure and scalability bottlenecks. They also provide limited verification and incentive mechanisms for mitigating unreliable or malicious agent behavior.



\subsection{Blockchain-Based Multi-Agent Systems}

Blockchain-based multi-agent systems offer decentralized mechanisms for orchestrating autonomous entities while providing immutable audit trails and programmable governance \cite{liu2025bewsat,liu2025bimw}.
DecoAgent \cite{jin2024decoagent} integrates LLM-powered agents with smart contracts to support decentralized autonomous collaboration, demonstrating how blockchain infrastructure can be used to coordinate agent interactions without relying exclusively on a centralized controller. 
BlockAgents \cite{chen2024blockagents} investigates Byzantine-robust LLM-based multi-agent coordination and uses blockchain mechanisms to improve the resilience and accountability of decentralized agent interactions.
LLM-Net~\cite{chong2025llm} designs blockchain-based expert networks to facilitate decentralized capability and resource sharing, while Qi~\textit{et al.}~\cite{qi2026towards} further examine transparent, incentive-compatible collaboration mechanisms among decentralized LLM agents.
Additionally, LLMChain~\cite{bouchiha2024llmchain} introduces a blockchain-based contextual reputation system that combines automated metrics and human feedback for decentralized LLMs management.

Nevertheless, these frameworks treat reputation and governance as passive tracking or settlement layers. They exhibit three primary limitations: (i) an inability to formalize complex, multi-stage task dependencies; (ii) static or decoupled agent selection policies; and (iii) the absence of closed-loop behavioral feedback in scheduling decisions. \textsc{Dart} addresses these shortcomings by integrating reputation-aware scheduling into active execution, using real-time behavioral evidence and DAG-modeled dependencies to drive adaptive, incentive-compatible multi-agent coordination.

\subsection{Incentive-Penalties and Reputation Mechanism}

Reputation and incentive schemes serve as foundational governance mechanisms in distributed systems, designed to foster sustained cooperation, enforce accountability, and suppress uncooperative behaviors.
In multi-agent environments, reputation provides a persistent representation of an agent's historical reliability, while incentives and penalties influence agents' decisions and participation.
With the emergence of LLM agents, recent research has integrated decentralized ledgers and smart contracts to establish verifiable, incentive-aligned multi-agent coordination.
For instance, BlockAgents \cite{chen2024blockagents} introduces Byzantine-robust coordination mechanisms to guarantee interaction integrity across autonomous entities.
DecoAgent \cite{jin2024decoagent} leverages smart contracts to facilitate decentralized agent governance and auditable workflow orchestration.
Furthermore, Qi et al. \cite{qi2026towards} explicitly investigate transparency and incentive compatibility in decentralized LLM multi-agent systems, demonstrating the importance of aligning individual agent behavior with system-level objectives.
These complementary behavior-driven studies have explored dynamic capability modeling, utility maximization, and workload balancing, confirming that incorporating historical performance profiles significantly outperforms static capability indexing.

Despite these advancements, existing approaches generally lack closed-loop integration among reputation, task allocation, workflow dependencies, and behavioral incentives. \textsc{Dart} addresses this gap by coupling reputation-aware selection, multi-factor incentives, and behavior-aware reputation updates within a unified regulation loop.



\section{\textsc{Dart} Framework Overview}
\label{sec:design}

\begin{figure}[t]
\centerline{\includegraphics[width=1\linewidth]{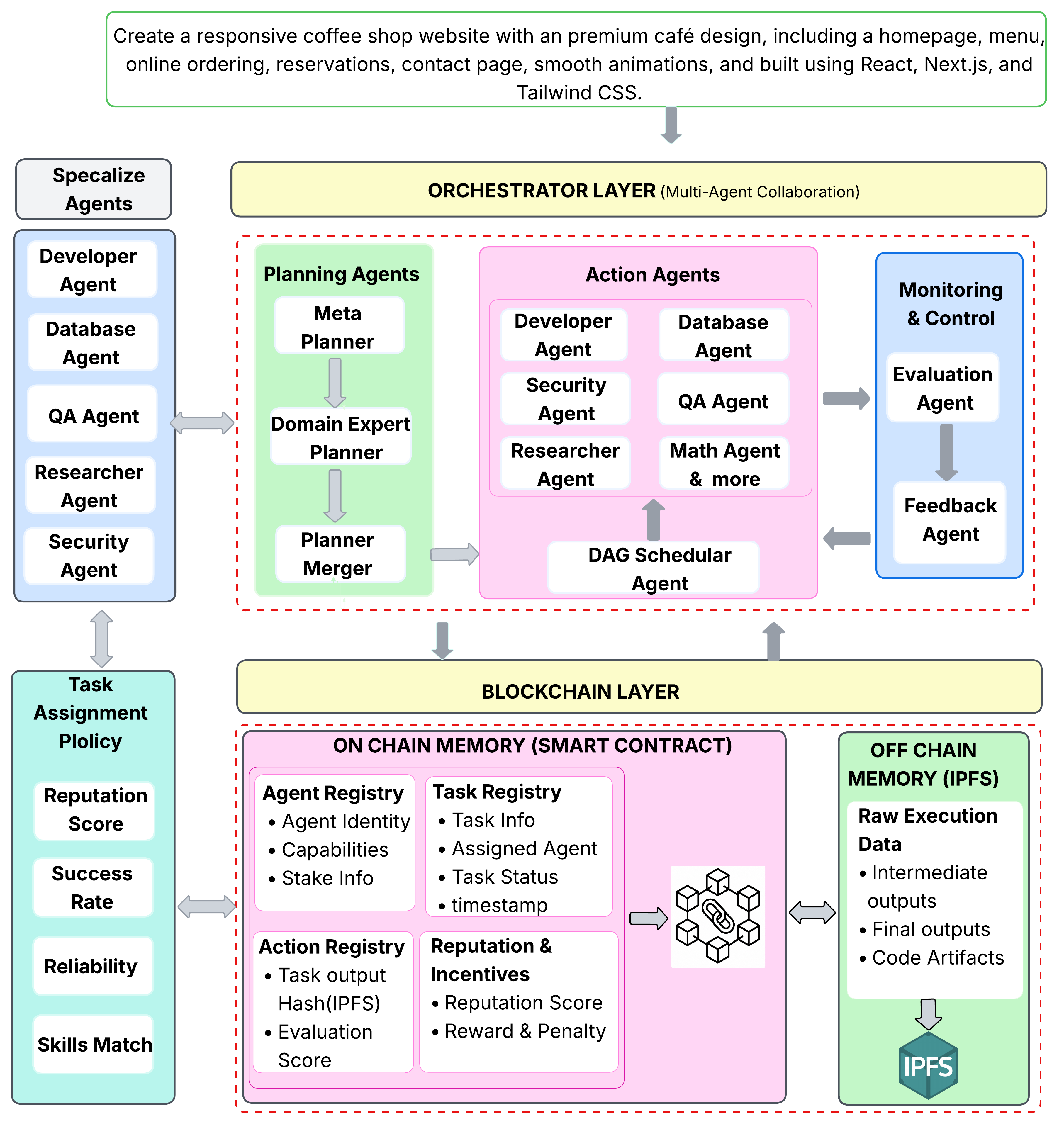}}
\caption{ System architecture of \textsc{Dart} for Decentralized LLM multi-agent collaboration with decentralized blockchain governance. }
\label{fig:dart_architecture}
\vspace{-10pt}
\end{figure}

In this section, we present \textsc{Dart}, a DAG-based agent reputation framework that fosters incentive-driven collaboration among  LLM-powered agents under decentralized blockchain governance. As depicted in Fig.~\ref{fig:dart_architecture}, the \textsc{Dart} architecture integrates specialized AI agents, policy-driven task allocation, a DAG-enabled orchestration pipeline, and a blockchain-based trust management system. This unified architecture enables 
 transparent and accountable task execution across heterogeneous agents while enforcing strict accountability via dynamic reputation, incentive, and penalty mechanisms.
The framework is structured into two core operational layers:
\begin{itemize}
    \item \textbf{Orchestrator Layer:} Executes the pipeline, including high-level task decomposition, capabilities-aware task allocation based on historical performance, output evaluation, and iterative feedback generation.
    
    \item \textbf{Blockchain Layer:} Enforces decentralized governance by managing agents and task registry, content-addressable storage, reputation scoring, incentive distribution, and immutable execution auditing.
\end{itemize}

DART adopts a hybrid architecture in which the orchestrator coordinates workflow planning and execution, while the blockchain layer provides decentralized governance, accountability, and immutable execution auditing. The workflow begins when a user submits a complex task requirement. The orchestrator decomposes it into an executable workflow and assigns subtasks based on agent capabilities. Generated outputs are evaluated, with failed outputs triggering corrective feedback for refinement or re-execution. State transitions, interaction traces, and evaluation metrics are recorded on-chain and off-chain to support auditing and future scheduling. This mechanism encourages cooperative coordination and collective problem-solving among specialized LLM agents.


\subsection{Orchestrator Layer}

The orchestrator layer manages the end-to-end lifecycle of task planning, execution, monitoring, and feedback through three core modules:

\textbf{1) Planning Module:} This module transforms high-level user requests into structured, executable subtasks via a three-stage planning pipeline:
\begin{itemize}
    \item \textit{Meta-Planner:} Parses user requests, performs high-level reasoning, and identifies overarching project objectives, operational constraints, task dependencies, and target deliverables. For instance, a web development request is decomposed into functional subtasks such as front-end design, back-end implementation, database schema design, security analysis, and quality assurance.
    
    \item \textit{Domain-Expert Planners:} Analyze decomposed subtasks to determine required domain-specific competencies and select specialized expert agents. These selected agents execute in parallel, independently proposing task-specific strategies tailored to their domain knowledge.
    
    \item \textit{Plan Merger:} Synthesizes the domain-specific proposals by eliminating redundant operations, resolving conflicting recommendations, and integrating optimal components into a unified, coherent execution plan.
\end{itemize}

\textbf{2) Action Module:} This module comprises the DAG Scheduler Agent and a pool of specialized execution agents. The DAG Scheduler Agent converts the consolidated plan into a Directed Acyclic Graph (DAG) representation to govern task dependencies, scheduling, and execution workflows~\cite{yang2026agentnet}. By mapping inter-task dependencies, the scheduler enforces correct execution sequences (e.g., ensuring front-end development initiates only after user interface specifications are finalized) while maximizing concurrency for independent tasks (e.g., executing database design and security analysis in parallel). 

Subtasks are dynamically allocated to candidate execution agents based on a composite suitability score derived from four key metrics: historical reputation, task success rate, operational reliability, and skill alignment. The candidate agent maximizing this suitability metric is assigned the task. The specialized agent pool encompasses diverse roles, including Developer Agents (code generation), Database Agents (schema optimization), Security Agents (vulnerability assessment), QA Agents (automated testing), and Researcher Agents (requirements analysis).

\textbf{3) Monitoring and Control Module:} This module enforces quality control over agent deliverables. An Evaluation Agent systematically assesses output completeness, correctness, and domain quality. If an output fails verification, a Feedback Agent synthesizes targeted corrective instructions and triggers an iterative re-execution loop with the assigned agent to refine the deliverable.

\subsection{Blockchain Layer}

The Blockchain layer establishes decentralized trust, transparency, and accountability across the agent ecosystem. By complementing centralized operational orchestration with smart-contract-based decentralized governance, this layer integrates a hybrid storage architecture combining on-chain computation with off-chain persistence to optimize storage efficiency and query scalability.

The on-chain memory layer relies on smart contracts to record critical system metadata, thereby avoiding the high gas costs associated with storing dense payload data directly on-chain. This lightweight state management encompasses four core registries:
\begin{itemize}
    \item \textbf{Agent Registry:} Manages agent onboarding and identity verification by storing cryptographic identities, registered capabilities, and collateral stake deposits.
    \item \textbf{Task Registry:} Maintains an immutable audit trail of task lifecycles, logging task specifications, agent assignments, execution statuses, and block timestamps.
    \item \textbf{Action Registry:} Captures operational outcomes by recording content hashes of generated deliverables alongside corresponding performance evaluation metrics.
    \item \textbf{Reputation and Incentive Registry:} Tracks dynamic reputation scores, reward distributions, penalty slashing events, and historical performance metrics to ensure transparent governance.
\end{itemize}

To prevent network congestion and prohibitive transaction costs, large execution artifacts such as intermediate reasoning steps, source code, documentation, and final deliverables are persisted in off-chain storage via the InterPlanetary File System (IPFS).
By storing only immutable and addressable IPFS content identifiers (CIDs) on-chain, the framework provides content-addressed integrity and verifiability while supporting low-latency data retrieval under the evaluated deployment conditions.

\section{Reputation-Based Regulation Mechanism}
\label{sec:reputation_incentive}

To promote sustainable cooperation and trustworthy coordination in blockchain-governed multi-agent environments, we formulate a behavior-shaping reputation and incentive mechanism (Algorithm~\ref{alg:reputation}).
The proposed framework dynamically regulates autonomous agent behavior through protocol-enforced incentives and penalties that:
\begin{enumerate}
    \item [(i)] Allocate economic rewards proportionally based on task complexity, LLM token efficiency, task execution success, and verified output quality;
    \item [(ii)] Impose penalties for capability mismatches, task failures, excessive execution latency, and behavioral anomalies; and
    \item [(iii)] Dynamically update long-term reputation scores by integrating instantaneous task performance with cumulative historical fidelity.
\end{enumerate}

\begin{algorithm}[t]
\caption{\textsc{Dart} Reputation Mechanism Workflow}
\label{alg:reputation}
\small
\begin{algorithmic}[1]
\Require Agent pool $\mathcal{A}$, task $T_j$, current state $\langle \rho_i^t, \mathbf{w}_i^t, L_i^t \rangle$ for each $A_i \in \mathcal{A}$, historical logs $\mathcal{H}_i^t$
\Ensure Selected agent $A_i^*$, updated state tuple $\langle \rho_{i^*}^{t+1}, L_{i^*}^{t+1} \rangle$, awarded incentive $I_{i^*,j}^t$

\vspace{2pt}
\PhaseComment{\textbf{Phase 1: Capability \& Reputation-Aware Allocation}}
\For{each agent $A_i \in \mathcal{A}$}
    \State Compute semantic capability match $M_{i,j}^t$ via \Comment{Eq.~\eqref{eq:capacity}}
    \State Compute composite suitability score: $\Phi_{i,j}^t$ \Comment{Eq.~\eqref{eq:sutability}}
\EndFor
\State Derive allocation distribution $\boldsymbol{\pi}_{\cdot,j}^t \leftarrow \operatorname{Softmax}(\mathbf{\Phi}_{\cdot,j}^t)$
\State Sample execution agent $A_i^* \sim \operatorname{Categorical}(\boldsymbol{\pi}_{\cdot,j}^t)$

\vspace{2pt}
\PhaseComment{\textbf{Phase 2: Task Execution \& Telemetry Collection}}
\State Assign $T_j$ to $A_i^*$ and record start time
\State Execute $T_j$ and collect telemetry: success $s_{i^*,j}^t$, quality $q_{i^*,j}^t$, latency $\tau_{i^*,j}^t$, token usage $u_{i^*,j}^t$
\State Perform specification verification to obtain $V_{i^*,j}^t \in [0,1]$

\vspace{2pt}
\PhaseComment{\textbf{Phase 3: Behavioral Anomaly \& Incentive Evaluation}}
\State Compute latency violation $D_{i^*,j}^t$ \Comment{Eq.~\eqref{eq:latency_violation}}
\State Compute behavioral anomaly score $B_{i^*,j}^t$ \Comment{Eq. \eqref{eq:behavioral_anomaly}}
\State Update sliding-window history $\mathcal{H}_{i^*, W}^t$ and evaluate windowed anomaly rate $\Psi_{i^*}^t$ \Comment{Eq. \eqref{eq:sliding_window}}
\State Compute positive reward $R_{i^*,j}^t$ and penalty $P^{\mathrm{e}}_{i^*,j}$ \Comment{Eq. \eqref{eq:rewarding}}
\State Award incentive value $I_{i^*,j}^t \leftarrow R_{i^*,j}^t - P^{\mathrm{e}}_{i^*,j}$ to agent $A_i^*$

\vspace{2pt}
\PhaseComment{\textbf{Phase 4: Multi-Factor Reputation \& Workload Update}}
\State Evaluate instantaneous performance score $P_{i^*,j}^t$ \Comment{Eq.~\eqref{eq:performance}}
\State Update cumulative historical failure rate $\Omega_{i^*}^t$ \Comment{Eq.~\eqref{eq:history_failure}}
\State Update long-term reputation score: $\rho_{i^*}^{t+1}$ \Comment{Eq.~\eqref{eq:reputation}}
\State Commit state $\langle \rho_{i^*}^{t+1}, L_{i^*}^{t+1} \rangle$ to smart contract for subsequent allocation rounds
\end{algorithmic}
\end{algorithm}

\subsection{System Modeling and State Representation}
Let $\mathcal{A} = \{A_1, A_2, \dots, A_n\}$ denote the set of $n$ autonomous agents, and $\mathcal{T} = \{T_1, T_2, \dots, T_m\}$ represent the set of $m$ candidate tasks. At time step $t$, the operational state of an agent $A_i \in \mathcal{A}$ is parameterized by a dynamic tuple $\langle \rho_i^t, \mathbf{w}_i^t, L_i^t \rangle$:
\begin{itemize}
    \item \emph{Reputation Score ($\rho_i^t \in [0,1]$):} Reflects the long-term reliability and historical execution fidelity of agent $A_i$.
    \item \emph{Capability Vector ($\mathbf{w}_i^t \in \mathbb{R}_{\ge 0}^d$):} Represents the expertise profile of agent $A_i$ across $d$ distinct skill domains.
    \item \emph{Workload Level ($L_i^t \in [0, 1]$):} Quantifies the normalized computational burden of agent $A_i$ relative to its maximum processing capacity $C_{i,\max}$:
    \begin{equation}
    \label{eq:capacity}
        L_i^t = \frac{1}{C_{i,\max}} \sum_{T_k \in Q_i^t} c_k,
    \end{equation}
    where $Q_i^t$ denotes the set of active subtasks assigned to $A_i$, and $c_k \in [0, 1]$ represents the normalized complexity score of subtask $T_k$.
\end{itemize}

Each task $T_j \in \mathcal{T}$ is characterized by a capability requirement vector $\mathbf{r}_j \in \mathbb{R}_{\ge 0}^d$, a normalized complexity score $c_j$, an execution deadline $d_j$, and a base reward structure.

\subsection{Utility-Based Suitability and Task Allocation}
The semantic alignment between an agent's capability profile $\mathbf{w}_i^t$ and task requirement $\mathbf{r}_j$ is evaluated via cosine similarity:
\begin{equation}
\label{eq:sutability}
M_{i,j}^{t} =
\begin{cases}
\dfrac{\mathbf{w}_{i}^{t} \cdot \mathbf{r}_{j}}
{\lVert\mathbf{w}_{i}^{t}\rVert_{2}\,\lVert\mathbf{r}_{j}\rVert_{2}},
& \text{if } \lVert\mathbf{w}_{i}^{t}\rVert_{2}\,\lVert\mathbf{r}_{j}\rVert_{2} > 0,\\[8pt]
0, & \text{otherwise}.
\end{cases}
\end{equation}

To assign task $T_j$, the orchestrator calculates a composite suitability score $\Phi_{i,j}^t$ that jointly accounts for capability matching, historical trustworthiness, and workload saturation:
\begin{equation}
\label{eq:workload}
    \Phi_{i,j}^t = \lambda_{1} M_{i,j}^t + \lambda_{2} (\rho_i^t \cdot M_{i,j}^t) - \lambda_{3} L_i^t,
\end{equation}
where $\lambda_1, \lambda_2, \lambda_3 $ are non-negative weighting coefficients satisfying $\sum_{k=1}^3 \lambda_k = 1$. The interaction term $(\rho_i^t \cdot M_{i,j}^t)$ prioritizes agents that possess both domain competence and demonstrated reliability, while the penalty term $-\lambda_3 L_i^t$ prevents structural task concentration and agent overloading.

To balance exploration and exploitation across the agent pool, the task assignment probability $\pi_{i,j}^t$ over all agents in $\mathcal{A}$ is derived via a Softmax transformation:
\begin{equation}
    \pi_{i,j}^t = \frac{\exp(\Phi_{i,j}^t)}{\sum_{A_k \in \mathcal{A}} \exp(\Phi_{k,j}^t)}.
\end{equation}
The target execution agent $A_i^* \sim \operatorname{Categorical}(\boldsymbol{\pi}_{\cdot,j}^t)$ is subsequently sampled from this distribution.


\subsection{Behavioral Anomaly and Verification Assessment}
Following the execution of task $T_j$ by agent $A_i$, the framework executes task-specific validation procedures (e.g., exact-answer comparison, test-case execution, constraint checking, or consistency verification against trusted evidence) to produce an objective verification score:
\begin{equation}
    V_{i,j}^t \in [0, 1],
    \label{eq:output verification}
\end{equation}
where $V_{i,j}^t = 1$ denotes complete compliance with the task specification.

To detect uncooperative, degraded, or adversarial behaviors, \textsc{Dart} aggregates multiple execution telemetry signals into an instantaneous behavioral anomaly score $B_{i,j}^t \in [0, 1]$:
\begin{equation}
\begin{aligned}
    B_{i,j}^t = {} & \omega_1 (1 - s_{i,j}^t) + \omega_2 (1 - q_{i,j}^t) + \omega_3 (1 - V_{i,j}^t) \\
                   & + \omega_4 (1 - M_{i,j}^t) + \omega_5 D_{i,j}^t, 
\end{aligned}
\label{eq:behavioral_anomaly}
\end{equation}
where $s_{i,j}^t \in \{0, 1\}$ is the binary execution success indicator, $q_{i,j}^t \in [0, 1]$ is the output quality score evaluated by \texttt{GPT-4o}, and $\sum_{r=1}^5 \omega_r = 1$ ($\omega_r \ge 0$). The term $D_{i,j}^t \in [0, 1]$ represents the normalized relative deadline violation:
\begin{equation}
\label{eq:latency_violation}
    D_{i,j}^t = \min\left\{1, \max\left[0, \frac{\tau_{i,j}^t - d_j}{d_j}\right]\right\},
\end{equation}
where $\tau_{i,j}^t$ denotes the observed execution latency.

To distinguish transient execution faults from systematic or intermittent malicious strategies, \textsc{Dart} maintains a short-term anomaly rate $\Psi_i^t$ evaluated over a sliding window of the $W$ most recent tasks assigned to $A_i$:
\begin{equation}
    \Psi_i^t = \frac{1}{|\mathcal{H}_{i, W}^t|} \sum_{\tau \in \mathcal{H}_{i, W}^t} B_{i,j(\tau)}^\tau,
    \label{eq:sliding_window}
\end{equation}
where $\mathcal{H}_{i, W}^t$ denotes the index set of the last $\min(|\mathcal{H}_i^t|, W)$ executed tasks by agent $A_i$.

\subsection{Incentive Formulation and Reputation Update}
The incentive value $I_{i,j}^t$ awarded to agent $A_i$ upon completing task $T_j$ balances baseline rewards against behavioral penalties:
\begin{equation}
    I_{i,j}^t = R_{i,j}^t - P^{\mathrm{e}}_{i,j},
    \label{eq:incentive}
\end{equation}
with the positive reward component $R_{i,j}^t$ and punitive deduction $P^{\mathrm{e}}_{i,j}$ defined as:
\begin{align}
    R_{i,j}^{t} &= \alpha_1 c_j
                + \alpha_2 s_{i,j}^{t}
                + \alpha_3 q_{i,j}^{t}
                + \alpha_4 u_{i,j}^{t},
    \label{eq:rewarding} \\
    P^{\mathrm{e}}_{i,j} &= \beta_1 (1 - M_{i,j}^t) + \beta_2 F_{i,j}^t + \beta_3 D_{i,j}^t + \beta_4 L_i^t + \beta_5 B_{i,j}^t,
    \label{eq:penalty}
\end{align}
where $\alpha_k, \beta_l \ge 0$ denote weighting hyperparameters, $u_{i,j}^t$ denotes the token consumption for executing task $T_j$ by agent $A_i$ , and $F_{i,j}^t = 1 - s_{i,j}^t$ denotes the instantaneous failure indicator. The normalized task complexity $c_j$ is defined over the DAG dependency depth $d_j^{\mathrm{dag}}$ and breadth $w_j^{\mathrm{dag}}$ as $c_j = \operatorname{norm}(d_j^{\mathrm{dag}} \cdot w_j^{\mathrm{dag}})$.

To update the long-term reputation profile, the instantaneous task performance $P_{i,j}^t$ is formulated as:
\begin{equation}
    P_{i,j}^t = \gamma_1 s_{i,j}^t + \gamma_2 q_{i,j}^t + \gamma_3 M_{i,j}^t + \gamma_4 V_{i,j}^t - \gamma_5 D_{i,j}^t,
    \label{eq:performance}
\end{equation}
where $\sum_{k=1}^5 \gamma_k = 1$ ($\gamma_k \ge 0$). 

Complementing the short-term anomaly rate $\Psi_i^t$, the cumulative historical failure rate $\Omega_i^t$ across all $|\mathcal{H}_i^t|$ historical tasks executed by $A_i$ is computed as:
\begin{equation}
\label{eq:history_failure}
    \Omega_i^t = \frac{1}{|\mathcal{H}_i^t|} \sum_{\tau \in \mathcal{H}_i^t} F_{i,j(\tau)}^\tau.
\end{equation}

Finally, the reputation score $\rho_i^{t+1}$ of agent $A_i$ is updated via an exponential moving average coupled with long-term penalty terms:
\begin{equation}
    \rho_i^{t+1} = \operatorname{clip}\left(\eta \rho_i^t + (1 - \eta)\left(P_{i,j}^t - \kappa \Omega_i^t - \mu \Psi_i^t\right), 0, 1\right),
    \label{eq:reputation}
\end{equation}
where $\eta \in [0, 1)$ acts as a memory retention coefficient, $\kappa \ge 0$ governs sensitivity to cumulative historical failures, and $\mu \ge 0$ enforces responsiveness to short-term behavioral anomalies.

\section{Experimental Results}
\label{sec:experiments}

To evaluate the effectiveness and robustness of \textsc{Dart}, we conduct comprehensive empirical assessments across both standard single-task benchmarks and complex multi-agent coordination scenarios.
While foundational reasoning and arithmetic benchmarks establish baseline competence, our primary evaluation targets complex tasks, like multi-stage software engineering and distributed social simulations, whose coordination overhead and execution complexity typically exceed the capabilities of isolated single LLM agents.
Furthermore, we systematically investigate the efficacy of \textsc{Dart}'s protocol-driven incentive formulations, verification-based reputation models, and anomaly-mitigation loops in governing blockchain-enabled multi-agent networks.
Our evaluation addresses four central Research Questions (RQs):
\begin{itemize}
    \item \textbf{RQ1 (Standard Benchmark Competence):} How effectively does \textsc{Dart} perform on standard symbolic, mathematical, and algorithmic benchmarks relative to state-of-the-art centralized and decentralized multi-agent baselines?
    \item \textbf{RQ2 (Hierarchical Scheduling \& Parallelism):} To what extent do \textsc{Dart}'s hierarchical multi-stage decomposition and DAG-based task scheduling streamline execution dependencies and maximize concurrent execution?
    \item \textbf{RQ3 (Incentive Alignment \& Accountability):} How do the smart-contract-enforced reputation dynamics and multi-factor incentive policies influence solution quality, resource efficiency, and agent accountability?
    \item \textbf{RQ4 (Anomaly Resilience):} How robustly does \textsc{Dart} detect, suppress, and isolate intermittent adversarial agent behaviors while preserving overall system performance and reliability?
\end{itemize}


\subsection{RQ1: Evaluation on Standard Benchmarks}
\label{sec:rq1_benchmarks}

\subsubsection{Experimental Setup}
Following established protocols \cite{zhang2025cut}, baseline reasoning and code generation capabilities are evaluated across four standard benchmarks: \emph{MBPP} \cite{austin2021program} (Python synthesis), \emph{HumanEval} \cite{chen2021evaluating} (docstring-to-code pass rates), \emph{MATH} \cite{hendrycks2021measuring} (competition mathematics), and \emph{GSM8K} \cite{cobbe2021training} (grade-school arithmetic). To ensure a fair comparison across MetaGPT, CAMEL, AgentVerse, AutoGen, MegaAgent, and DART, all frameworks use the same \texttt{GPT-4o} backend with a temperature of 0 and identical datasets and evaluation prompts. Performance is measured by using top-1 pass accuracy ($\text{Pass}@1$, \%)

\subsubsection{Results and Comparative Analysis}

\begin{table}[ht]
\centering
\caption{Performance Comparison ($\text{Pass}@1$ Accuracy, \%) Across Standard Code Generation and Mathematical Reasoning Benchmarks.}
\label{tab:Multiagent_framework_results_comparison}
\setlength{\tabcolsep}{9pt}
\begin{tabular}{lcccc}
\toprule
\textbf{Framework} & \textbf{MBPP} & \textbf{HumanEval} & \textbf{MATH} & \textbf{GSM8K} \\
\midrule
MetaGPT    & 81.7 & 82.3 & --   & --   \\
CAMEL      & 78.1 & 57.9 & 22.3 & 45.6 \\
AgentVerse & 82.4 & 89.0 & 54.5 & 81.2 \\
AutoGen    & 85.3 & 85.9 & \textbf{69.5} & 87.8 \\
MegaAgent  & \textbf{92.2} & \textbf{93.3} & 69.0 & 93.0 \\
\midrule
\textsc{Dart} (Ours) & 81.0 & 88.0 & 64.0 & \textbf{93.6} \\
\bottomrule
\end{tabular}
\end{table}

Table~\ref{tab:Multiagent_framework_results_comparison} summarizes the comparative performance across all four benchmarks.
As reported in Table~\ref{tab:Multiagent_framework_results_comparison}, \textsc{Dart} exhibits robust general-purpose problem-solving competence, achieving the highest overall accuracy on GSM8K ($93.6\%$), outperforming both MegaAgent ($93.0\%$) and AutoGen ($87.8\%$). This highlights the effectiveness of our dynamic task-decomposition and capability-matching mechanism in mathematical reasoning and multi-step problem-solving tasks.

On software synthesis benchmarks (MBPP and HumanEval), \textsc{Dart} attains competitive accuracies of $81.0\%$ and $88.0\%$, respectively, matching or exceeding several general-purpose frameworks (e.g., CAMEL and MetaGPT on HumanEval). While specialized coding architectures such as MegaAgent achieve higher pass rates ($92.2\%$ on MBPP and $93.3\%$ on HumanEval) due to domain-specific prompt engineering and iterative code-compilation loops, \textsc{Dart} maintains balanced, cross-domain multi-agent orchestration without domain-tailored heuristics.
Similarly, on the rigorous MATH dataset, \textsc{Dart} scores $64.0\%$, demonstrating solid multi-step algebraic deduction.
These baseline results demonstrate that the hybrid orchestration and blockchain-enabled governance mechanisms introduced in \textsc{Dart} do not compromise underlying core problem-solving fidelity.

\subsection{RQ2: Complex Multi-Agent Coordination via Gobang Game Development}
\label{sec:rq2_gobang}
To evaluate \textsc{Dart’s} ability to manage complex workflows requiring multi-role planning, functional autonomy, and parallel execution, we use the development of the two-player strategy game Gobang (Gomoku) as a multi-agent software engineering benchmark. The task requires heterogeneous agents, including software architects, GUI/UX designers, game programmers, and integration testers, to concurrently develop game logic, event handling, AI move generation, and graphical rendering.


\subsubsection{Experiment Setup}

\begin{figure}[htbp]
\centerline{\includegraphics[width=1\linewidth]{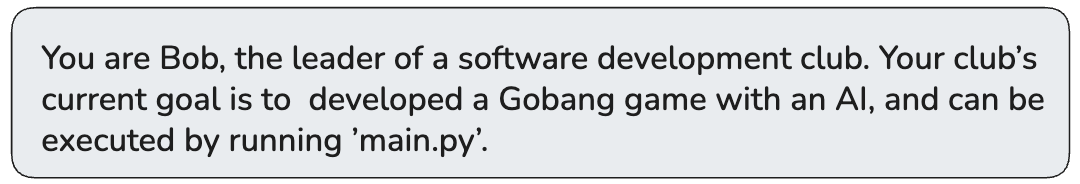}}
\caption{Meta-prompt specification utilized for the end-to-end Gobang game development scenario.}
\label{fig:meta-prompt}
\end{figure}

All frameworks are evaluated using identical prompts and LLM configurations with GPT-4o via Azure OpenAI at $\tau=0$. One legacy baseline that does not support GPT-4o uses GPT-4 instead. Frameworks are initialized with the structured Gobang meta-prompt, while baseline configurations follow their original SOPs and recommended settings. We compare DART with AutoGen, MetaGPT, CAMEL, AgentVerse, and MegaAgent.
Framework initialization is guided by the structured meta-prompt detailed in Fig.~\ref{fig:meta-prompt}.

\subsubsection{Evaluation Criteria}
We assess the functional completeness and operational robustness of the synthesized applications against four qualitative criteria:
\begin{itemize}
    \item \textbf{C1 (Error-Free Execution):} Evaluates whether the delivered codebase compiles and runs cleanly via the target entry point without unhandled runtime exceptions or syntax crashes;
    \item \textbf{C2 (User Interaction):} Verifies whether the GUI correctly registers user mouse clicks, validates board boundaries, and updates board states without input lag;
    \item \textbf{C3 (AI Opponent Generation):} Assesses whether the system synthesizes a functional automated opponent capable of generating valid, non-colliding response moves;
    \item \textbf{C4 (Termination Verification):} Assesses whether the game engine accurately detects five-in-a-row winning patterns across horizontal, vertical, and diagonal orientations, terminates gameplay, and announces the victor.
\end{itemize}

Additionally, we capture quantitative computational overheads: total number of orchestrated agents, cumulative wall-clock execution time (seconds), and normalized time per agent.

\subsubsection{Experimental Results and Comparative Analysis}
Table~\ref{tab:gameresults} presents the functional validation and efficiency benchmarks across all evaluated platforms.
As shown in Table~\ref{tab:gameresults}, \textsc{Dart} and MegaAgent are the only frameworks that successfully fulfill all four functional criteria (\textbf{C1}--\textbf{C4}).
However, \textsc{Dart} demonstrates substantial efficiency gains, delivering a fully operational application in only $142\,\text{s}$ using an optimized 2-agent DAG-orchestrated pipeline, compared to $800\,\text{s}$ across 7 agents in MegaAgent (an $82.2\%$ reduction in latency).

\begin{table}[htbp]
\caption{Gobang Game Development Results}
\label{tab:gameresults}
\centering
\setlength{\tabcolsep}{4pt}
\begin{tabular}{lccccccc}
\toprule
\textbf{Framework} &
\makecell[c]{\textbf{C1}} &
\makecell[c]{\textbf{C2}} &
\makecell[c]{\textbf{C3}} &
\makecell[c]{\textbf{C4}} &
\makecell[c]{\textbf{\#Agents}} &
\makecell[c]{\textbf{Time(s)}} &
\makecell[c]{\textbf{Time/Agent (s)}}\\
\midrule

AutoGen
& {\color{Green}\cmark}
& {\color{Green}\cmark}
& {\color{red}\xmark}
& {\color{red}\xmark}
& 2
& 120
& 60 \\

MetaGPT
& {\color{Green}\cmark}
& {\color{Green}\cmark}
& {\color{red}\xmark}
& {\color{red}\xmark}
& 6
& 480
& 80 \\

CAMEL
& {\color{red}\xmark}
& {\color{red}\xmark}
& {\color{red}\xmark}
& {\color{red}\xmark}
& 2
& 1,168
& 584 \\

AgentVerse
& {\color{red}\xmark}
& {\color{red}\xmark}
& {\color{red}\xmark}
& {\color{red}\xmark}
& 4
& 1,980
& 495 \\

MegaAgent
& {\color{Green}\cmark}
& {\color{Green}\cmark}
& {\color{Green}\cmark}
& {\color{Green}\cmark}
& 7
& 800
& 114 \\

DART (Our)
& {\color{Green}\cmark}
& {\color{Green}\cmark}
& {\color{Green}\cmark}
& {\color{Green}\cmark}
& 2
& \textbf{142}
& \textbf{71} \\
\bottomrule

\end{tabular}
\end{table}

A detailed diagnosis of baseline failure modes reveals key orchestration bottlenecks:
\begin{itemize}
    \item \emph{AutoGen:} Operates with a two-agent coder--executor conversational loop. Although it produces a runnable interface within $120\,\text{s}$, it omits complex AI heuristics and win-condition validations (\textbf{C3}, \textbf{C4}) due to the lack of explicit task decomposition and automated code-review stages.
    \item \emph{MetaGPT:} Deploys a 6-agent waterfall architecture. While generating compliant boilerplate UI code, it suffers from context fragmentation and lacks iterative functional debugging, failing to integrate dynamic opponent logic (\textbf{C3}) and termination checks (\textbf{C4}) within $480\,\text{s}$.
    \item \emph{CAMEL:} Employs unconstrained dual-agent role-playing. It suffers from conversational drift and inadequate context retention, resulting in non-functional scripts that fail to execute basic syntax (\textbf{C1}) despite running for $1,168\,\text{s}$.
    \item \emph{AgentVerse:} Utilizes a 4-agent collaborative consensus loop. Over multiple trials, it encounters recursive inter-agent rejection cycles and generates fragmented code containing unpopulated placeholders, failing all verification criteria after $1,980\,\text{s}$.
\end{itemize}

\subsubsection{Ablation and Resource Cost Analysis}
To investigate the token utilization efficiency of \textsc{Dart}'s hierarchical orchestration, we evaluate resource consumption across three operational stages: \textit{Hierarchical Planning}, \textit{Parallel Task Solving}, and \textit{Artifact Merging}.
Table~\ref{tab:token_usage} reports the corresponding token overheads and phase durations.
The empirical breakdown provides two critical insights:

\textbf{Concurrent Pipeline Overlap:} The transition between Planning ($0 \rightarrow 45\,\text{s}$) and Task-Solving ($30 \rightarrow 140\,\text{s}$) exhibits temporal concurrency, demonstrating that our DAG scheduler dispatches independent sub-modules immediately upon resolution without blocking on the complete plan.

\textbf{Controlled Token Overhead:} By structuring contextual boundaries through dedicated subtask prompts, the Task-Solving stage generates full backend logic and UI rendering with only $47,887$ total tokens, preventing the quadratic context explosion typical of unstructured multi-agent dialogue loops.

\begin{table}[htbp]
\caption{Stage-Wise Token Consumption and Execution Latency in \textsc{Dart}}
\label{tab:token_usage}
\centering
\setlength{\tabcolsep}{10pt}
\begin{tabular}{lrrrr}
\toprule
\textbf{Stage} & \textbf{Input} & \textbf{Output} & \textbf{Total} & \textbf{Time (s)} \\
\midrule
Planning      & 2,529  & 1,867 & 4,396  & 0-45 \\
Task-Solving  & 40,307 & 7,580 & 47,887 & 30-140 \\
Merging       & 18,723 & 5,656 & 24,379 & 140-160 \\
\midrule
\textbf{Total} & \textbf{61,559} & \textbf{15,103} & \textbf{76,662} & \textbf{160} \\
\bottomrule
\end{tabular}
\end{table}

\subsection{RQ3: Incentive Alignment and System Trustworthiness}
\label{sec:rq3_trustworthiness}

To evaluate the operational impact of \textsc{Dart}'s reputation, incentive, and penalty mechanisms, we systematically investigate their capacity to shape agent behavior and foster overall system trustworthiness. 
Specifically, we evaluate whether reputation-guided scheduling drives long-term convergence toward high task completion rates, superior output quality, optimal capability--task alignment, and stable workload distribution across heterogeneous agent groups.

\subsubsection{Experimental Setup}

\begin{table}[tbp]
\caption{Taxonomy of Agent Capability Tags for Profile Vectors ($\mathbf{w}_i$)}
\label{tab:capability_taxonomy}
\centering
\small
\begin{tabular}{ll}
\toprule
\textbf{Tag} & \textbf{Description} \\
\midrule
Tag1  & Object recognition and classification \\
Tag2  & Spatial reasoning and planning \\
Tag3  & Language understanding (instruction parsing) \\
Tag4  & Grasping and manipulation \\
Tag5  & Path planning and navigation \\
Tag6  & Scene understanding (layout/context extraction) \\
Tag7  & Task decomposition and sequencing \\
Tag8  & Temporal reasoning (event ordering, not deadlines) \\
Tag9  & Knowledge grounding (external inference) \\
Tag10 & Environment interaction via API calls or actuators \\
\bottomrule
\end{tabular}
\end{table}

Our experimental environment is constructed upon the tool-augmented function-calling suite of AgentBench \cite{liu2024agentbench}.
The benchmark corpus comprises 150 diverse multi-step tasks spanning five representative operational domains: ALFWorld (embodied decision-making), DBBench (relational database querying), KnowledgeGraph (structured knowledge reasoning), OS Interaction (operating system shell execution), and ALFRED (interactive household task synthesis).
Tasks encompass both atomic functional invocations and complex multi-stage workflows.

Task requirements are formalized using the 10-dimensional capability taxonomy outlined in Table~\ref{tab:capability_taxonomy}.
To isolate reputation dynamics from difficulty-induced variance, the 150 tasks are stratified across three difficulty tiers (\textit{Easy}, \textit{Medium}, \textit{Hard}).
Workloads are grouped into 150-round evaluation epochs, each maintaining an identical $1:1:1$ distribution across all three difficulty levels.
We instantiate $19$ heterogeneous autonomous agents, each characterized by distinct initial capability vectors $\mathbf{w}_i^0$. All agents are initialized with a neutral longitudinal reputation score ($\rho_i^0 = 0.50$).

To improve statistical robustness and quantify stochastic variation, each configuration was evaluated over five independent simulation runs using distinct random seeds (30, 32, 42, 50, and 60).
Each run consisted of 150 sequential rounds, with each round representing a complete lifecycle of task arrival, task allocation, execution profiling, independent verification, incentive distribution, and state updating.
This configuration was evaluated over five independent 150-round trials.
To systematically isolate the contributions of capability matching, reputation tracking, workload balancing, incentive regulation, and anomaly detection, we evaluate six ablation variants: 
(i)~Capability Only (baseline), (ii)~+~Reputation, (iii)~+~Workload, (iv)~Full \textsc{Dart} $-$ Incentive, (v)~Full \textsc{Dart} $-$ Anomaly, and (vi)~Full \textsc{Dart}.


\subsubsection{Results and Comparative Analysis}









\begin{table*}[t]
\centering
\small
\caption{Ablation Study of \textsc{Dart} Components for Task Allocation and Execution}
\label{tab:ablation_study}
\resizebox{\textwidth}{!}{
\begin{tabular}{l*{5}{c}*{4}{r}}
\toprule
\textbf{Method} &
\textbf{Cap.} &
\textbf{Rep.} &
\textbf{Work.} &
\textbf{Inc.} &
\makecell[c]{\textbf{Anom.}\\\textbf{Det.}} &
\makecell[c]{\textbf{Success}\\\textbf{Rate (\%) $\uparrow$}} &
\makecell[c]{\textbf{Quality}\\\textbf{Score $\uparrow$}} &
\makecell[c]{\textbf{Task Retry}\\\textbf{Rate (\%) $\downarrow$}} &
\makecell[c]{\textbf{Allocation}\\\textbf{Delay $\downarrow$}} \\
\midrule

Capability Only
& \checkmark & $\times$ & $\times$ & $\times$ & $\times$
& 89.07 $\pm$ 2.14 & 0.9144 $\pm$ 0.0098 & 0.3480 $\pm$ 0.0693 & 1.1740 $\pm$ 0.0347 \\

+ Reputation
& \checkmark & \checkmark & $\times$ & $\times$ & $\times$
& 90.40 $\pm$ 1.92 & 0.9195 $\pm$ 0.0092 & 0.3120 $\pm$ 0.0357 & 1.1560 $\pm$ 0.0179 \\

+ Workload
& \checkmark & \checkmark & \checkmark & $\times$ & $\times$
& 90.93 $\pm$ 3.08 & 0.9253 $\pm$ 0.0158 & 0.2987 $\pm$ 0.0732 & 1.1493 $\pm$ 0.0366 \\

\textsc{Dart} $-$ Incentive
& \checkmark & \checkmark & \checkmark & $\times$ & \checkmark
& 90.10 $\pm$ 1.53 & 0.9220 $\pm$ 0.0088 & 0.3080 $\pm$ 0.0540 & 1.1540 $\pm$ 0.0270 \\

\textsc{Dart} $-$ Anomaly Detection
& \checkmark & \checkmark & \checkmark & \checkmark & $\times$
& 91.60 $\pm$ 2.48 & 0.9257 $\pm$ 0.0135 & 0.2760 $\pm$ 0.0651 & 1.1380 $\pm$ 0.0325 \\

\textbf{Full \textsc{Dart}}
& \checkmark & \checkmark & \checkmark & \checkmark & \checkmark
& \textbf{93.33 $\pm$ 2.26} & \textbf{0.9357 $\pm$ 0.0117} & \textbf{0.2307 $\pm$ 0.0816} & \textbf{1.1153 $\pm$ 0.0408} \\

\bottomrule
\multicolumn{10}{l}{\parbox{\linewidth}{\vspace{2pt}\footnotesize\textit{Note:} Values are reported as mean $\pm$ standard deviation across five independent 150-round runs using random seeds 30, 32, 42, 50, and 60. $\uparrow$ indicates higher is better; $\downarrow$ indicates lower is better.}}
\end{tabular}
}
\vspace{-10pt}
\end{table*}

    



Table \ref{tab:ablation_study} presents the ablation results across five independent 150-round simulation runs.
All results are reported as the arithmetic mean and standard deviation ($\text{mean} \pm \text{SD}$), capturing both average system performance and variability across random seeds. 


\paragraph{Task Execution Fidelity and Failure Suppression} 
Full \textsc{Dart} achieves the highest mean task success rate (\textbf{93.33 $\pm$ 2.26\%}), outperforming the baseline Capability Only configuration (\textbf{89.07 $\pm$ 2.14\%}) and intermediate variants.
Integrating reputation and workload awareness yields progressive performance gains, whereas omitting incentive regulation or anomaly detection degrades overall execution efficacy. 
These results demonstrate that combining reputation, workload, incentive, and anomaly mechanisms maximizes task execution performance.
Additionally, Full \textsc{Dart} records the lowest mean task retry rate (\textbf{0.2307 $\pm$ 0.0816}), significantly lower than Capability Only \textbf{0.3480 $\pm$ 0.0693}.
This reduction demonstrates that reputation and workload-aware allocation enhances initial execution reliability, while incentive and anomaly mechanisms enforce robust behavioral regulation.


\paragraph{Solution Quality and Outcome Consistency} 
Full \textsc{Dart} achieves the highest mean output quality (\textbf{0.9357 $\pm$ 0.0117}), outperforming the Capability Only baseline (\textbf{0.9144 $\pm$ 0.0098}).
The progressive improvement with reputation and workload integration,
alongside performance drops when omitting incentive or anomaly detection mechanisms,
demonstrate that the complete regulation loop improves output quality while maintaining low variation across independent runs.

\paragraph{Incentive and Reputation Dynamics} 
The ablation results indicate that the individual components provide complementary improvements rather than a single mechanism accounting for the overall gain.
Introducing reputation-aware allocation improves task success and output quality over the Capability Only baseline, while workload balancing further improves result quality and reduces retry frequency.
Removing incentive regulation decreases success rate and increases retry frequency relative to Full \textsc{Dart}, indicating that incentive regulation contributes to sustained execution reliability.
Similarly, removing anomaly detection increases the retry rate from \textbf{0.2307} to \textbf{0.2760} and decreases the mean success rate from \textbf{93.33\%} to \textbf{91.60\%},
demonstrating that behavioral regulation is essential for suppressing unreliable executions.

\paragraph{Scheduling Overhead and Workload Balance} 

Full \textsc{Dart} achieves the lowest mean allocation delay (\textbf{1.1153 $\pm$ 0.0408~s}), compared with \textbf{1.1740 $\pm$ 0.0347~s} for Capability Only.
The progressive reduction with reputation and workload awareness, along with the higher delays when incentive or anomaly detection is removed, indicates that the complete \textsc{Dart} configuration provides the most efficient task allocation.

\subsubsection{Domain-Specific Reputation Dynamics}





\begin{figure}[htbp]
\centering
\includegraphics[width=0.95\linewidth]{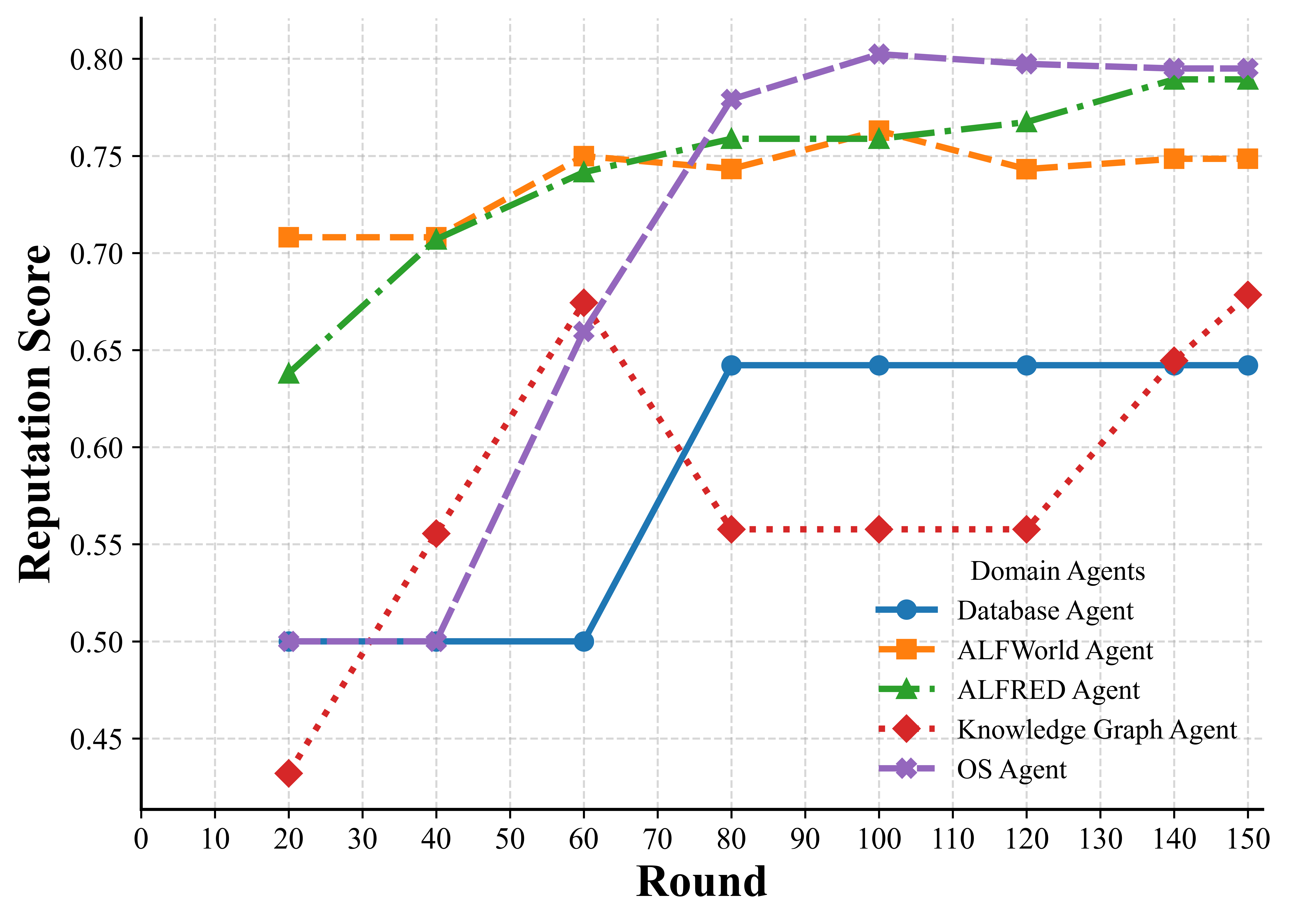}
\vspace{-4pt} 
\caption{Agent reputation across domains.}
\label{fig:reputation}
\vspace{-10pt} 
\end{figure}

Fig.~\ref{fig:reputation} illustrates the trajectory of longitudinal reputation scores ($\rho_i^t$) across distinct agent domains alongside workload variance over the 150-round simulation.
The reputation trajectories demonstrate \textsc{Dart}'s ability to dynamically update trust profiles in response to continuous execution feedback.
\begin{itemize}
    \item \emph{Rapid Trust Accumulation:} Agents operating in structured, highly deterministic domains (e.g., OS Interaction) exhibit rapid reputation growth, advancing from $\rho_i^0 = 0.50$ to $0.80$ before plateauing, reflecting consistent rule-based execution.
    \item \emph{Sustained High Performance:} Embodied agents (e.g., ALFWorld, ALFRED) converge smoothly within the $0.75\text{--}0.79$ interval, establishing a high trust baseline through sustained manipulation and navigation accuracy.
    \item \emph{Resilient Trust Recovery:} Agents handling complex reasoning tasks (e.g., KnowledgeGraph) encounter occasional execution drops due to domain ambiguity, resulting in temporary reputation dips. However, \textsc{Dart}'s sliding-window update rule permits subsequent successful executions to rehabilitate trust scores, indicating that the governance framework distinguishes between intermittent execution variance and systemic unreliability.
\end{itemize}

\subsection{RQ4: Adversarial Robustness and Anomaly Isolation}
\label{sec:rq4_adversarial_robustness}

In real-world multi-agent deployments, participating agents cannot be assumed to be inherently trustworthy. Capable adversaries may generate stealthy or corrupted outputs while strategically maintaining high capability scores to evade immediate heuristic checks. To evaluate \textsc{Dart}'s capacity to identify and contain such threats, we evaluate its multi-layered security mechanics: independent verification Eq.~\eqref{eq:output verification}, behavioral anomaly scoring Eq.~\eqref{eq:behavioral_anomaly}, sliding-window anomaly aggregation Eq.~\eqref{eq:sliding_window}, and anomaly-aware reputation updates Eq.~\eqref{eq:reputation}.

\subsubsection{Experiment Setup} 

We construct a controlled adversarial environment (simulation) over 150 simulation rounds.
The agent pool consists of the 19 honest agents evaluated in Section~\ref{sec:rq3_trustworthiness}, augmented with three programmatically injected malicious agents ($N = 22$).
System configurations, task distributions, and capability profiles are consistent with the settings of the RQ3 evaluation to isolate the effects of adversarial behavior.
We instantiate three distinct adversarial personas:
\begin{itemize}
    \item \emph{Agent Beta (Persistent Byzantine Adversary):} Corrupts $100\%$ of assigned task outputs, representing a persistent worst-case attack vector.
    \item \emph{Agent Alpha (Intermittent Strategic Adversary):} Corrupts $70\%$ of assigned task outputs while executing standard, honest operations for the remaining $30\%$. This models a strategic attacker that attempts to evade detection by maintaining a baseline level of positive execution history.
    \item \emph{Agent Gamma (Moderately Intermittent Adversary):} Corrupts 60\% of assigned task outputs while producing honest executions for the remaining 40\%. Gamma represents a more stealthy adversary that introduces a larger proportion of legitimate behavior than Alpha, thereby reducing the short-term anomaly signal and increasing the difficulty of timely isolation
\end{itemize}

All three adversaries are initialized with elevated capability profiles and high nominal quality ratings to simulate capable, high-reputation attackers: Alpha ($\mathbf{w}_{\text{match}} = 0.91$, $q = 0.92$), Beta ($\mathbf{w}_{\text{match}} = 0.93$, $q = 0.94$) and Gamma ($\mathbf{w}_{\text{match}} = 0.89$, $q = 0.90$). To allow adversaries to establish initial trust within the system, malicious behavior is activated at Round 30 ($t_{\text{attack}} = 30$).

For every completed task, \textsc{Dart} independently verifies the output and computes the behavioral anomaly score $B_{i,j}^t$ using the weighting vector $\boldsymbol{\omega} = (0.30, 0.25, 0.25, 0.10, 0.10)$ across task failure, output quality, verification score, capability mismatch, and execution latency, respectively. Suspicious execution evidence is aggregated over a sliding window of $W = 5$ recent tasks. An agent is flagged and permanently quarantined from the task-allocation pool when its short-term anomaly rate satisfies $\Psi_i^t \ge \theta_{\text{flag}} = 0.65$.

\subsubsection{Evaluation Metrics}
We assess system robustness across seven complementary dimensions:
\begin{itemize}
    \item \textbf{Detection Time (DT):} The latency (in rounds) between attack activation ($t = 30$) and formal agent isolation;
    \item \textbf{Anomaly Flag Rate (AR):} The percentage of malicious executions whose instant anomaly score exceeds the detection threshold;
    \item \textbf{Malicious Reputation Trajectory (MR):} The temporal decay of longitudinal reputation scores ($\rho_i^t$) for adversarial agents;
    \item \textbf{Malicious Selection Rate (MSR):} The proportion of total system tasks allocated to malicious agents across successive evaluation windows;
    \item \textbf{False Positive Rate (FPR):} The percentage of benign agents incorrectly flagged and quarantined;
    \item \textbf{Output Containment Rate (OCR):} The proportion of corrupted outputs successfully blocked from downstream workflow integration;
    \item \textbf{System-Level Resiliency:} Overall task success rate, average solution quality, and failure rate across three distinct operational phases (\textit{Pre-Attack}, \textit{Active Attack}, and \textit{Post-Isolation}).
\end{itemize}

\subsubsection{Experimental Results and Robustness Analysis}

\begin{figure}[htbp]
\centerline{\includegraphics[width=1\linewidth]{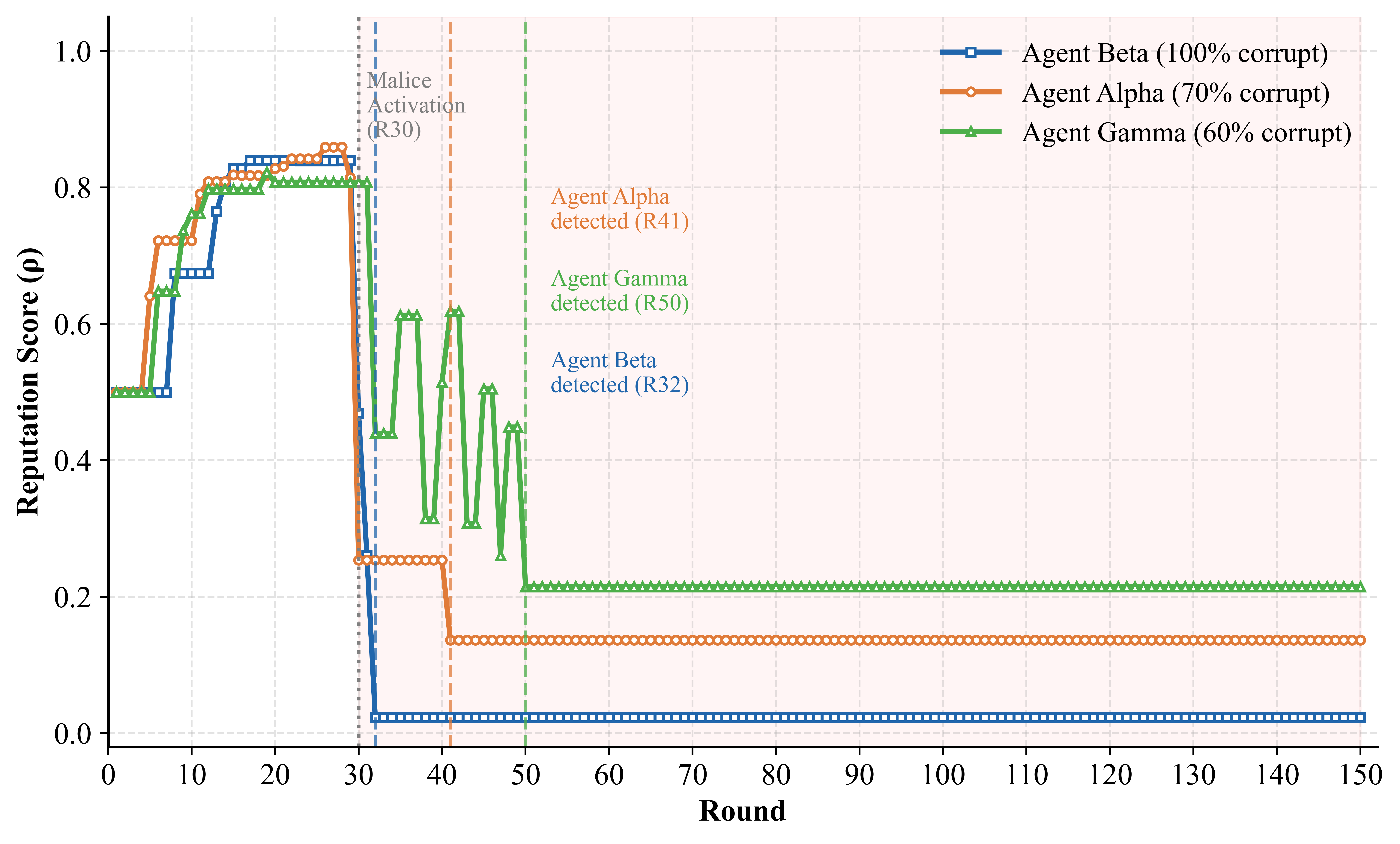}}
\caption{Longitudinal reputation trajectories ($\rho_i^t$) of persistent and intermittent (Alpha and Gamma) adversarial agents, highlighting attack activation ($t = 30$) and quarantine isolation milestones.}
\label{fig:anomaly agent reputation}
\vspace{-10pt}
\end{figure}

\textsc{Dart} successfully identified and isolated all three malicious agents without incurring any false positives ($\text{FPR} = 0\%$) in the evaluated adversarial simulation.
Fig.~\ref{fig:anomaly agent reputation} tracks the reputation degradation and isolation points of the adversarial agents.
The persistent adversary (Beta) was flagged and quarantined at Round 32, yielding a detection time of $\text{DT}_{\text{Beta}} = 2$ rounds.
In contrast, the intermittent adversary (Alpha) was isolated at Round 41 $\text{DT}_{\text{Alpha}} = 11$ rounds.
Similarly, the moderately intermittent adversary (Gamma) was isolated at Round 50, yielding $\text{DT}_{\text{Gamma}}=20$ rounds.
The extended detection windows for Alpha and Gamma reflect their strategic insertion of honest executions, which temporarily dampened their sliding-window anomaly average ($\Psi_{\text{Alpha}}^t$) and delayed quarantine threshold crossing.

The overall task-level anomaly flag rate was $21.9\%$, capturing instant execution anomalies.
Meanwhile, \textsc{Dart}'s allocation policy rapidly suppressed task dispatch to adversarial nodes via continuous stages:
\begin{itemize}
    \item \emph{Rounds 1--50 (Initialization \& Early Attack):} Malicious agents captured $12.4\%$ of total task allocations while building and leveraging initial trust.
    \item \emph{Rounds 51--100 (Isolation Phase):} Following Beta, Alpha, and Gamma's progressive removal at Rounds 32, 41, and 50, MSR dropped precipitously to $0.0\%$.
    \item \emph{Rounds 101--150 (Post-Isolation Phase):} Following the quarantine of all adversarial agents, MSR remained at $0.0\%$, eliminating adversarial task participation entirely.
\end{itemize}

Reputation updates under Eq.~\eqref{eq:reputation} further demonstrate the effectiveness of the penalty term $-\mu \Psi_i^t$.
Beta's reputation collapsed rapidly from $\rho_{\text{Beta}}^{30} = 0.50$ to $0.0570$ by Round 32,
while Alpha’s and Gamma’s reputations declined more gradually due to their intermittent honest executions. Furthermore, \textsc{Dart} achieved an overall $\text{OCR} = 99.3\%$, successfully intercepting corrupted artifacts during the active attack phase prior to quarantine.

\begin{table}[t]
\caption{System Resiliency Across Operational Phases Under Adversarial Attack}
\label{tab:phase_performance}
\centering
\small
\setlength{\tabcolsep}{4pt} 
\begin{tabular}{lcccc}
\toprule
\textbf{Operational Phase} & \textbf{Rounds} & \makecell[c]{\textbf{Success}\\\textbf{Rate (\%)}} & \makecell[c]{\textbf{Mean}\\\textbf{Quality}} & \makecell[c]{\textbf{Failure}\\\textbf{Rate (\%)}} \\
\midrule
Pre-Attack Baseline & 1--29   & \textbf{100.0} & \textbf{0.85} & \textbf{0.0} \\
Active Attack       & 30--64  & 87.6           & 0.74          & 12.4          \\
Post-Isolation      & 65--150 & 99.8           & 0.84          & 0.2          \\
\bottomrule
\end{tabular}
\vspace{-10pt}
\end{table}

Table~\ref{tab:phase_performance} summarizes system performance across the three operational phases.
During the active attack window (Rounds 30--64), system success rate temporarily decreased to $87.6\%$ and output quality dropped to $0.74$. However, following the isolation of all three malicious agents, system performance recovered rapidly during Rounds 65--150, restoring the task success rate to $99.8\%$ and solution quality to $0.84$, matching pre-attack baseline levels. 
These empirical results address RQ4 affirmatively: \textsc{Dart} effectively isolates persistent and intermittent adversaries without benign false positives, contains over $99\%$ of corrupt outputs, and restores overall multi-agent system performance to near-ideal baseline fidelity.

\subsection{Blockchain and Storage Overhead Analysis}
\label{sec:blockchain_storage_overhead}

\begin{table}[t]
\caption{System Latency and Gas Consumption Across On-Chain and Off-Chain Operations}
\label{tab:system_performance}
\centering
\small
\renewcommand{\arraystretch}{1.15}
\setlength{\tabcolsep}{4pt}
\begin{tabular}{llcc}
\toprule
\textbf{Layer} & \textbf{Operation} & \makecell[c]{\textbf{Mean Latency}\\\textbf{(ms)}} & \makecell[c]{\textbf{Gas}\\\textbf{Used}} \\
\midrule
\multirow{4}{*}{\makecell[l]{On-Chain\\(EVM)}} 
 & Agent Registration   & 4.40 & 166,078 \\
 & Task Registration    & 6.70 & 168,802 \\
 & Reputation Update    & 5.50 &  47,974 \\
 & Incentive Settlement & 4.90 &  29,235 \\
\midrule
\multirow{2}{*}{\makecell[l]{Off-Chain\\(IPFS)}} 
 & Artifact Upload      & 31.70 & -- \\
 & Artifact Retrieval   & 2.59  & -- \\
\bottomrule
\end{tabular}
\vspace{-10pt}
\end{table}

To evaluate \textsc{Dart}'s operational overhead, we benchmark on-chain smart contract state transitions and off-chain IPFS artifact management. On-chain benchmarking measures execution latency and gas costs across agent registration, task instantiation, reputation updates, and incentive settlement on a local Hardhat Ethereum node ($12\,\text{M}$ gas limit). Off-chain profiling measures upload/retrieval latencies for execution logs ($\le 20\,\text{kB}$).
As detailed in Table~\ref{tab:system_performance}, \textsc{Dart}'s hybrid architecture effectively decouples blockchain state verification from large-scale artifact storage overhead.
On-chain operations incur transaction gas fees proportional to the amount of smart contract state modified in EVM storage.
\textit{Agent Registration} ($166,078$ gas) and \textit{Task Registration} ($168,802$ gas) incur higher gas costs due to the initialization of persistent, multi-attribute dynamic data structures in smart contract storage.
In contrast, \textit{Reputation Updates} ($47,974$ gas) and \textit{Incentive Settlement} ($29,235$ gas) require less gas because they modify existing fixed-size states.
Off-chain IPFS operations consume zero blockchain gas while maintaining sub-second performance. The asymmetric latency profile between artifact upload ($31.70\,\text{ms}$) and retrieval ($2.59\,\text{ms}$) demonstrates that
offloading execution traces to IPFS prevents state bloat on-chain, ensuring that the results indicate the feasibility of the proposed hybrid architecture under the evaluated local deployment conditions.


\section{Discussions and Limitations}
\label{sec:discussion}
\subsection{Discussion of Key Findings}

\textsc{Dart} maintains competitive performance on standard benchmarks (\textbf{RQ1}) while supporting multi-stage multi-agent coordination. In \textbf{RQ2}, DAG-based execution reduced the reported application completion time relative to the centralized baseline. In \textbf{RQ3}, the five-run ablation study showed that Full \textsc{Dart} achieved the strongest overall performance across task success, output quality, retry rate, and allocation delay. In \textbf{RQ4}, \textsc{Dart} isolated the evaluated persistent and intermittent adversaries while maintaining a low false-positive rate and recovering system performance after isolation. Finally, local EVM/IPFS measurements demonstrate the feasibility of the hybrid implementation, while public-network scalability remains future work.

\subsection{Parameter Sensitivity and Hyperparameter Calibration}

To assess \textsc{DART's} robustness to hyperparameter choices, we conduct a controlled sensitivity analysis of the principal parameters governing agent allocation and behavioral anomaly detection. We vary the anomaly sliding-window size $W$, detection threshold $\theta_{\mathrm{detect}}$, and allocation weights $(\lambda_1,\lambda_2,\lambda_3)$ while fixing all other parameters at their default values. Each configuration uses a synthetic simulation with the same agent population and task structure as the main pipeline. Results are averaged over \(N\) independent seeds to isolate hyperparameter effects from LLM-inference variance.
The results reveal predictable parameter trade-offs rather than dependence on a narrowly tuned configuration. Increasing $W$ increases detection delay from 6.1 rounds at $W=3$ to 48.0 rounds at $W=20$, reflecting the trade-off between rapid detection and stable reputation estimates. Similarly, $\theta_{\mathrm{detect}}=0.35$ reduces detection delay to 7.3 rounds but yields a 13.2\% false-quarantine rate, whereas $\theta_{\mathrm{detect}}\geq0.55$ eliminates false quarantines at the cost of slower detection, reaching 15.7 rounds. In contrast, allocation performance is comparatively insensitive to the exact $(\lambda_1,\lambda_2,\lambda_3)$ allocation, with composite scores across the evaluated configurations spanning less than 1.5\% of the observed range; balanced weighting performs marginally best.
Overall, these results indicate that DART is comparatively robust to allocation-weight choices while exhibiting predictable detection responsiveness to $W$ and $\theta_{\mathrm{detect}}$. Although the evaluated ranges do not indicate dependence on a single narrowly tuned configuration, optimal settings may vary across workloads and threat models, motivating adaptive calibration as a direction for future work.

\subsection{Limitations of Adversarial Evaluation}

Our adversarial evaluation considers three canonical attack profiles: one persistent adversary with 100\% output corruption and two intermittent adversaries with 70\% and 60\% corruption.
These profiles validate \textsc{Dart's} baseline containment mechanism but do not capture fully adaptive, threshold-aware adversaries.
Specifically, an attacker with knowledge of the reputation dynamics could regulate its attack frequency or interleave malicious and benign executions to remain below the quarantine threshold and evade detection.
Evaluating \textsc{Dart} against such adaptive adversaries remains an important direction for future work.



\section{Conclusions}
\label{sec:conclusion}
This paper presented \textsc{Dart}, a hybrid DAG-based agent reputation and incentive framework designed to establish trustworthy and accountable collaboration in LLM-based multi-agent systems.
\textsc{Dart} addresses trust and accountability limitations associated with centralized orchestration by combining DAG-based workflow coordination with blockchain-enabled decentralized governance. While operational planning, scheduling, task allocation, and execution coordination remain managed by the orchestrator, the blockchain layer provides tamper-resistant reputation management, incentive settlement, and immutable execution auditing.
To ensure scalability and transparency, the architecture decouples lightweight governance state managed on-chain via smart contracts from heavy execution artifacts stored in content-addressable IPFS networks. 
Empirical evaluations demonstrate that \textsc{Dart} effectively preserves task quality and enforces long-term system integrity without compromising execution efficiency.

\bibliographystyle{IEEEtranS}
\bibliography{reference}

\end{document}